\documentclass[10pt,letterpaper]{article}
\usepackage{opex3}
\usepackage{cite}

\begin{document}

\title{3D mapping of intensity field about the focus of a micrometer--scale parabolic mirror}

\author{Alison McDonald$^1$, Gail McConnell$^1$, David C. Cox$^2$, Erling Riis$^3$, and Paul F. Griffin$^{3,*}$}
\address{$^1$Strathclyde Institute of Pharmacy and Biomedical Sciences, University of Strathclyde, Glasgow, G4 0RE, UK\\
$^2$Advanced Technology Institute, University of Surrey, Guildford, Surrey GU2 7XH, UK\\
$^3$Department of Physics, SUPA, University of Strathclyde, Glasgow G4 0NG, UK}

%

\email{$^*$paul.griffin@strath.ac.uk} 



\begin{abstract}
We report on the fabrication and diffraction-limited characterization of parabolic focusing micromirrors. Sub--micron beam waists are measured for mirrors with 10-$\mu$m radius aperture and measured fixed focal lengths in the range from 24~$\mu$m to $36~\mu$m. Optical characterization of the 3D intensity in the near--field produced when the device is illuminated with collimated light is performed using a modified confocal microscope. Results are compared directly with angular spectrum simulations, yielding strong agreement between experiment and theory, and identifying the competition between diffraction and focusing in the regime probed. 
\end{abstract}



\section{Introduction}
The introduction of reflective optical elements revolutionized the design and imaging capabilities of telescopes \cite{King:book}. While the possibility of creating large diameter objectives was one of the primary aims, a significant advantage is that reflective optical elements provide a means of reducing or eliminating the spherical and chromatic aberrations that are inherent in lens--based systems. 

Here we explore the opposite extreme of microfabricated optics. Microfabricated optical elements are key components for the development and integration of optics into a range of research and commercial areas \cite{NovotnyHecht:book,JamrozKruzeleckyHaddad:book,Voegel:MicroEng:2003}. To date, the majority of the work in microphotonics has been in refractive elements, i.e. microlenses. However, in this regime microlenses typically have significant numerical apertures and surface curvatures, which introduce large aberrations. 

A number of groups have in recent years discussed the design and fabrication of concave micromirrors \cite{Trupke:APL:2005,Salathe:OptExpr:2007,Salathe:OptLett:2009, Azimi:OptExpr:2010, Blain:AppPhysLett:2010}. These examinations are largely driven by two purposes; for optical tweezing, and for integration into atom optics. Spherical mirrors have been demonstrated to collect light from single ions \cite{Blinov:PRA:2010,Slusher:NJP:2011}. Parabolic mirrors, similar to those described here but with larger length scales, have been also been used as highly efficient collectors of light from single ions \cite{Leuchs:PRA:2012, Leuchs:AppPhysB:2014}, atoms \cite{Barrett:NJP:2012}, and point sources \cite{Kim:OptLett:2010}. In a similar manner they may be used to tightly focus light onto atomic samples, which has to date been shown with refractive optics \cite{Kurtsiefer:NatPhys:2008}.

In this work we consider the use of reflective micro--optical components for focusing light. We present the construction and optical characterization of parabolic reflectors with an open aperture of radius 10~$\mu$m and measured focal lengths that range from 24~$\mu$m to 36~$\mu$m. Detailed mapping of the focused intensity field is made possible by the development of a previously--unreported adaptation of a confocal microscope that allows the illumination of the reflector with collimated light, while still maintaining the highly-desirable large-numerical-aperture confocal collection. Using this device we obtain 3D data about the focal plane demonstrating diffraction limited focussing. We also discuss the application of the parabolic mirror for use in atomic physics and tweezing experiments.


\section{Fabrication}
The details of the fabrication of concave paraboloid structures through ion-beam milling are covered in \cite{Cox:2014}. Briefly, a focused ion-beam (gallium ions with typical currents of 50-300~pA and accelerating voltages of 30~kV) is used to precisely sculpt a silicon substrate with the required mirror profile, which is subsequently coated with gold to provide a highly-reflective coating. In focused ion-beam milling, controlling the dose of ions to which an area is exposed allows a region of the surface to be sputtered to a known depth, due to a linear relationship between depth and dose in silicon. The applied dose is a function of the beam current, the dwell time, and number of passes the beam makes over an area. By tracing a number of concentric discs of increasing radius, whilst linearly increasing the dose, a parabolic depression can be milled into the substrate. In principle this would create a stepped contour, however due to edge-effects of the milling process, as well as redistribution of etched material, a larger number of passes creates a smooth contour of the parabolic dish. Further details of the construction and characterization can be found in \cite{Cox:2014}, where an RMS roughness of 4.0~nm was measured by AFM over the range of the concave parabolic surface. Due to the identical manufacturing process similar values are expected in this work.


\section{Theory}
The propagation of light using microscopic optical elements is a well represented topic in the literature. Here we restrict our discussion to the behavior of light fields after wavelength--scale apertures and curved surfaces. In their work, Goldwin and Hinds \cite{Hinds:OptExpr:2008} derive analytic results for spherical mirrors, which they further compare with numerical integration of Maxwell's equations. Meanwhile, Bandi {\it et. al.} consider the propagation of light after a wavelength-scale aperture using a Fresnel representation of the fields \cite{NicChormaic:PRA:2008}, which offers the possibility of adding focusing to the formalism. 

\begin{figure}[b!]
\centerline{\includegraphics[width=9cm]{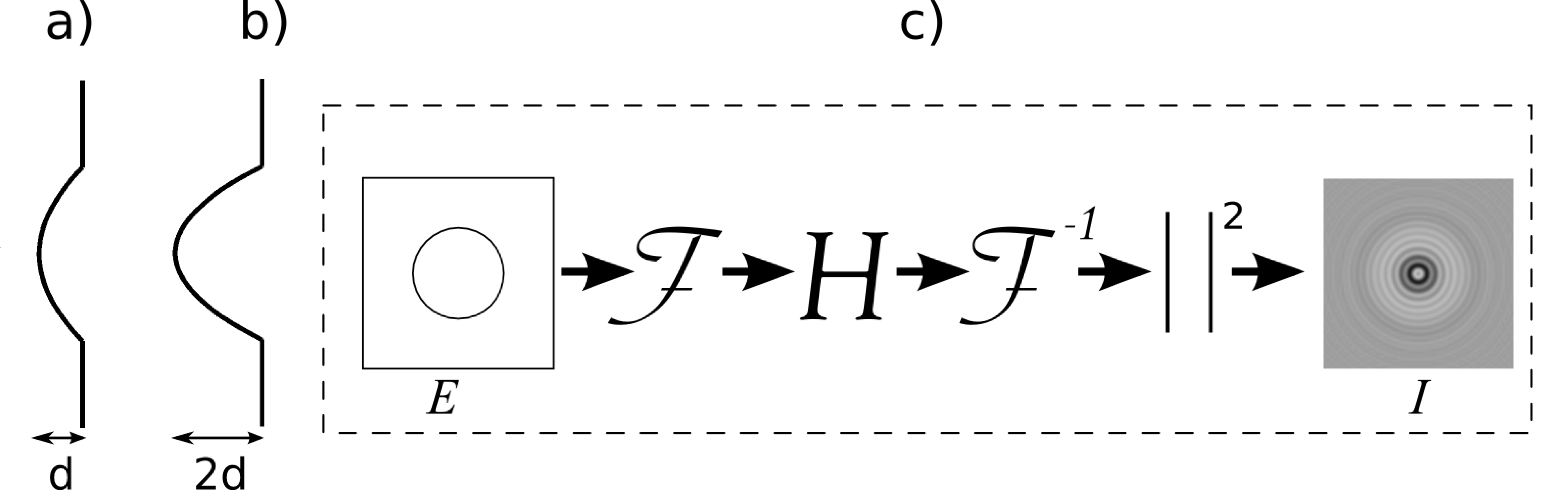}}
\caption{Details of the theoretical modeling of the micro-mirrors. The depth-profile of the mirror (a) causes a phase delay (b) equivalent to a optical path length of twice the depth of the mirror. The intensity at an arbitrary plane due to this phase delay are then found using the angular spectrum method (c), the details of which are in the text.}
\label{fig:theory_schematic}
\end{figure}

Here, however, the reflected field was modeled using the angular spectrum method \cite{NovotnyHecht:book, Goodman:book}, which provides a mapping of an electric field from a particular plane into a secondary plane; 
\begin{equation}
{\bf E}(x,y,z) = \int\limits_{-\infty}^{\phantom{}\infty}\int \limits_{-\infty}^{\phantom{}\infty}  \hat{\bf E}(k_x,k_y;z\!=\!0)\,
{\rm e}^{i\left (k_x x + k_y y \pm k_z z\right)} {\rm d}k_x \rm{d}k_y~,
\end{equation}
where $x,y,z$ are the Cartesian spatial directions, $k_i$ is the wavevector in the $i$-direction, ${\bf E}$ is the electric field, and $\hat{\bf E}(k_x,k_y;z)$ is the 2D Fourier spectrum of the electric field in the plane $z$. The result of this compact equation can be understood by first noting that the spatial spectrum of an electric field, $\hat{\bf E}$ can be translated from a plane $z=0$ to another plane $z$ using the Helmholtz propagator, 
\begin{equation}
\hat{\bf E}(k_x,k_y;z)\, = \hat{\bf E}(k_x,k_y;z\!=\!0)\,{\rm e}^{\pm k_z \,z}~,
\end{equation}
where the Helmholtz propagator in reciprocal space is $\hat{H}=\exp(\pm k_z \,z)$, \cite{NovotnyHecht:book}. We then note that the electric field can be calculated from its spatial spectrum in a plane are by the inverse Fourier transform, ${\bf E}(x,y,z) = \mathcal{F}^{-1}\hat{\bf E}(k_x,k_y;z)$,
\begin{equation}
{\bf E}(x,y,z) = \int\limits_{-\infty}^{\phantom{}\infty}\int \limits_{-\infty}^{\phantom{}\infty}  
\hat{\bf E}(k_x,k_y;z) {\rm e}^{i\left(k_x x + k_y y\right)} {\rm d}k_x {\rm d}k_y
\end{equation}
These relations, Eq.~(2) and Eq.~(3) then clearly show the result of Eq.~(1), and can be used to calculate the electric field in an arbitrary plane, given that is is known in one plane. The essential details of calculating the final intensity profile using the angular spectrum method can be clearly seen in Fig.~\ref{fig:theory_schematic}(c); the initial electric field is Fourier transformed, the Helmholtz propagator is then applied, before the inverse Fourier spectrum is taken. Finally, the resulting intensity field is found from the modulus-squared of the electric field. Example codes are available on request from the corresponding author.

For this work, in the plane of $z=0$ a curvature is numerically added to the phase-fronts within the area of the aperture of the mirror. The curvature is expressed as $\exp\left(2\,i\, k\, d(r)\right)$, where $k=2\,\pi/\lambda$ is the total wavevector, $d(r)=r^2/4f$, $r<a$ is the radial distance from the center of the mirror, $a$ is the mirror aperture radius, and $f$ is the focal length. This modification of the optical wavefront represents the position dependent optical path-length difference, due to the spatially varying propagation distance to the metal surface, Fig.~\ref{fig:theory_schematic}. Care must be taken to include all modes of the field, including the evanescent ones, in order to return the full field simulation. All simulations were performed in MATLAB with the focal length, $f$ and the position of the surface as free parameters for fitting to the measured data.


\section{Probing}

\begin{figure}[b!]
\centerline{\includegraphics[width=9cm]{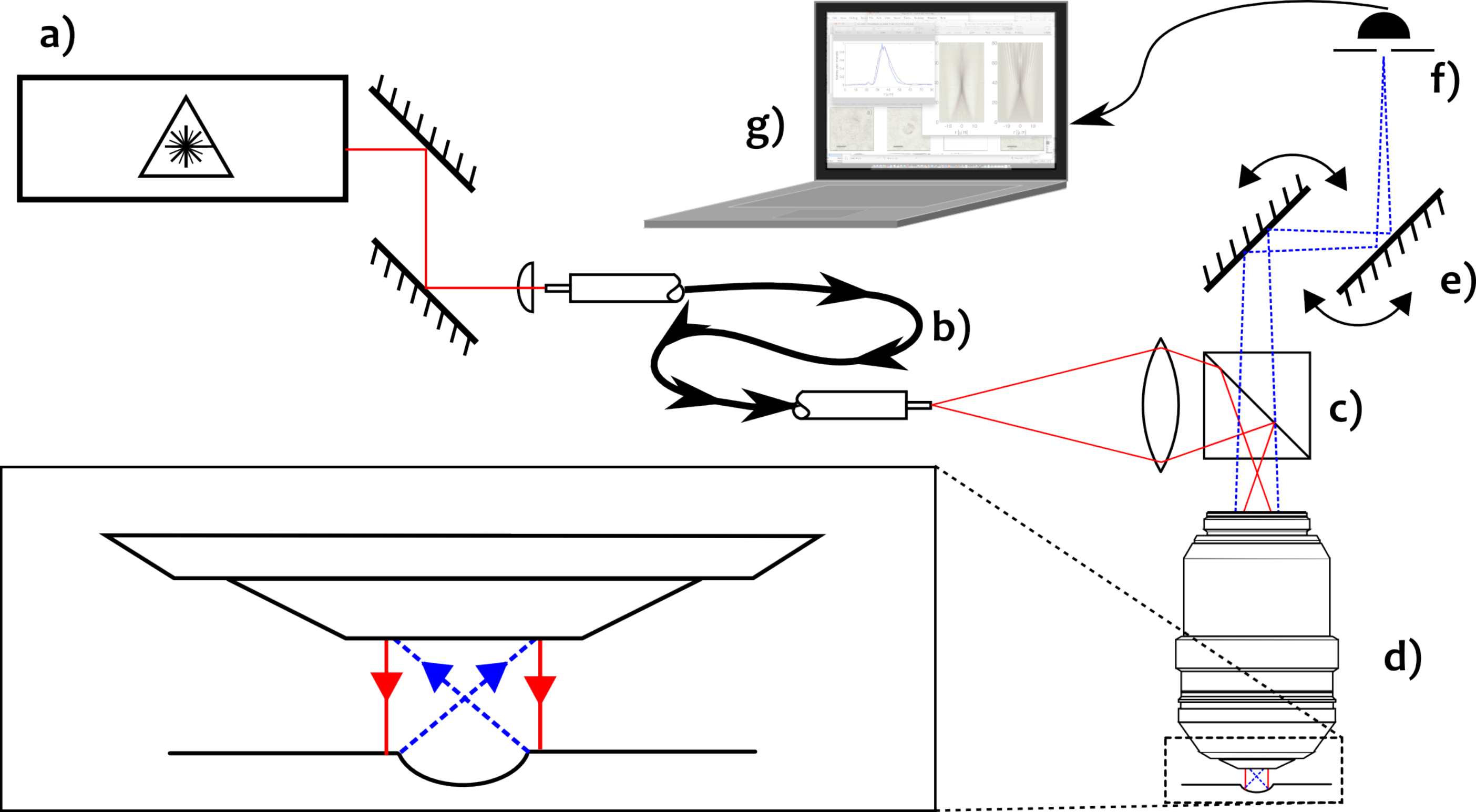}}
\caption{The optical layout used to probe the parabolic mirrors in infinite conjugation. A 589~nm laser (a) is fiber-coupled (b) to a non-polarizing beam splitter (c) and mode-matched through the rear of the microscope objective (d). The image was rastered within the microscope (e) onto the confocal pinhole (f) and the data compiled on a computer (g). Inset; showing the collimated output of the microscope. Solid (red) lines after the optical fiber indicate the illumination path, blue (dashed) lines the imaging path.}
\label{fig:confocal_schematic}
\end{figure}

Although transmission confocal imaging provides a powerful means of examining microlenses \cite{Dawson:2004:APL:a}, reflection images of focusing elements require post analysis to interpret the data. In traditional reflection confocal microscopy the illuminating laser is focused onto the imaging plane at a point that is confocal with a pinhole in the imaging axis of the microscope. While this allows accurate probing of the surface it precludes imaging the focal plane of our concave mirrors directly. Instead one may directly image about the center of curvature of the mirror and then infer the transfer function of the mirror assuming the point spread functions of the other optical elements are known. In this work we devise a scheme that allows interrogation of the focus while illuminating a mirror with a collimated laser beam, Fig.~\ref{fig:confocal_schematic}. This arrangement means that the intensity distribution that would be formed in an optical tweezing experiment \cite{Ashkin:1992:BiophysJ:a}, for example, can be faithfully interrogated.

Probing the focal plane of the parabolic mirrors requires illumination with collimated light co--axial with the mirror axis (orthogonal to the substrate surface). The working distance of the objective lens used (Olympus UPlanSApo 40x/0.9$NA$) is only 180~$\mu$m, which severely limits introducing further optical elements before the object. To circumvent these restrictions a non-polarizing beam splitter (NPBS) was introduced between the objective lens and the scanning column of a commercial confocal microscope (Olympus FV1000 scan head with Olympus IX81 inverted microscope). This enabled an external probe beam to be introduced through the back aperture of the objective lens using the NPBS to combine the beam into the imaging axis, see Fig.~\ref{fig:confocal_schematic}. The probe, a commercial solid-state 589~nm laser (based on sum-frequency generation using two lasing lines of an Nd:YAG laser), was focused, after fiber-coupling to clean up the spatial mode, at the back-focal plane of the objective lens. The specific probe wavelength was chosen for convenience (good detector efficiency and high optical power in a single transverse mode) - the numerical simulations were also conducted for this wavelength to account for wavelength-dependent diffraction effects. The collimation of the beam emitted was optimized by varying the axial position of the coupling lens. The NPBS, optical fiber, and coupling lens were mounted in a holder with three--axis adjustment, which allowed for overlapping of the optical axes. The emitted beam had a diameter of 1~mm, justifying the assumption of a spatially uniform intensity profile across the micromirror aperture (20~$\mu$m diameter). The magnification of the combined optical system was calibrated by measuring the aperture diameter of the mirrors.

In order to visually verify the surface quality of the mirrors, as well as to measure the aperture diameter of the mirrors the surface of the mirrors was also probed using conventional confocal microscopy \cite{Dawson:2004:APL:a}, without removal of the additional optical elements. No aberrations were observed to have been introduced.


\section{Results}

\begin{figure}[b!]
\centerline{\includegraphics[width=8.5cm]{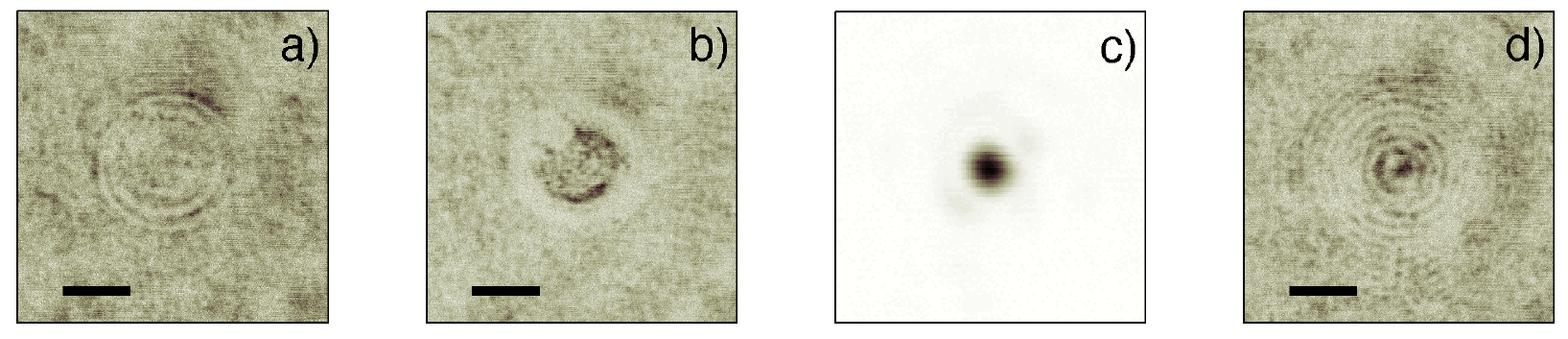}}
\caption{Raw data from modified confocal microscope for a $f=36~\mu$m parabolic micromirror. a) shows data collected from the surface of the mirror substrate; b) taken 25~$\mu$m before the focus ($z=11~\mu$m in Fig~\ref{fig:data_vs_simulation}), c) taken at the position of the focus, and d) taken 25~$\mu$m after the focus ($z=61~\mu$m in Fig~\ref{fig:data_vs_simulation}). The black scale bars in a), b), d) indicate a length of 10~$\mu$m, which is the width of image c). The linear colormap for each image is normalized from white to black for minimum to maximum value respectively.}
\label{fig:raw_data}
\end{figure}
Optical measurements were taken for four mirrors with different focal lengths, all with 10~$\mu$m radii apertures. Two dimensional scans were taken, with 77~nm pixel size, of the reflected intensity field in defined planes parallel to the surface of the substrate, Fig.~\ref{fig:raw_data}. A series of such scans, separated by 1~$\mu$m, then return a high-resolution, 3D map of the intensity field.

\begin{figure}[h!]
\centerline{\includegraphics[width=12cm]{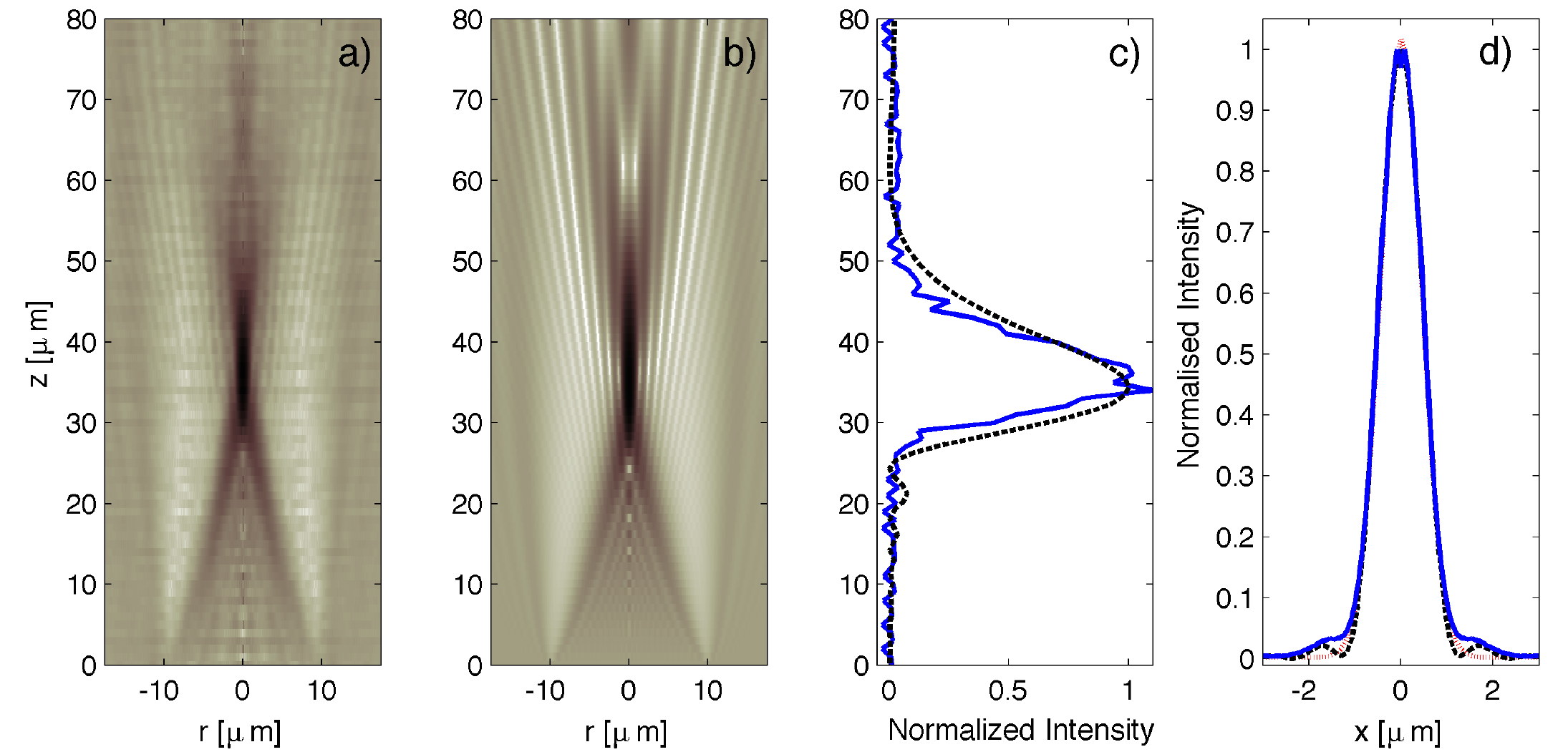}}
\caption{Comparison of the measured a) and simulated data b) from the $f=36~\mu$m mirror, using a log-scale to highlight the detail in the intensity field away from the focus. The data in each figure are each normalized, with the same color-axis for both. The color-axis, [-4 0], corresponds to a range of four orders of magnitude in intensity. The measured (solid blue line) and simulated (dashed line) on-axis intensity profiles, c), and radial intensity profiles at the focus, d), show agreement between experiment and model. Also included in d), as a red dotted line, is a Gaussian fit to the simulated data.}
\label{fig:data_vs_simulation}
\end{figure}

The raw data were then radially averaged about the optical axis of the beam. The optical axis was independently found by fitting a 2D Gaussian to individual scans near the focal plane and then linearly extrapolating the center--of--mass fits to regions where a simple 2D-Gaussian fit was not valid, {\it e.g.}, such as in Fig.~\ref{fig:raw_data}(b). (Note, however, that about the focus, Fig.~\ref{fig:raw_data}(c), a 2D-Gaussian gives an excellent fit to the data.) This step was taken to ensure any residual tilts in the optical system were accounted for. The averaged data were then combined, as shown in Fig.~\ref{fig:data_vs_simulation}(a), to visualize the cylindrically symmetric intensity field. The focal length was extracted from the distance between the surface and the peak of the intensity, as well as by the best fit of the numerical model, Fig.~\ref{fig:data_vs_simulation}(b). These data show excellent qualitative agreement as the detailed structure of the field away from the focus, which is emphasized by the logarithmic scale on the color axis. To further illustrate the agreement between experiment and theory the on-axis intensity profile of the data and the simulation are shown in Fig.~\ref{fig:data_vs_simulation}(c). 

Quantitative results from the measurements are in Table~\ref{Table}, below. The measured waists given are the minimum e$^{-2}$ radii returned from Gaussian fits about the waist. The waists in the orthogonal directions were found to be equal and to occur in the same place, within measurement error. It can be seen in Fig.~\ref{fig:data_vs_simulation}(d) that a Gaussian is an excellent fit to both simulation and experimental data near the focus.
The diffraction limit was calculated as $\lambda/2\,{N\!A}$, where ${N\!A}$ is the numerical aperture, $N\!A=\sin(2\,\tan^{-1}(a/f)$  which was derived using the measured focal length \cite{Salathe:OptExpr:2007}. It is clear that there is good agreement between the measured minimum beam waists and the prediction from diffraction theory. 

\begin{table}[h!]
  \caption{Measured beam waists (e$^{-2}$ radius) and focal lengths for a range of parabolic mirrors with 10~$\mu$m radius aperture. Errors, where quoted, are one standard deviation. }
\label{Table}
  \begin{center}
    \begin{tabular}{ccccc}
    \hline
    Mirror  & Measured	& Measured		& Simulated\\
	       & {\it f}  [$\mu$m] & waist [nm] 		& Waist [nm]	\\
    \hline
    1 & 36(1) 	& 875(22)	&	886(3)\\
    2 & 27.5(5)	& 736(26)  	&	673(4)\\
    3 & 25.5(5)	& 765(23) 	&	627(3)\\
    4 & 24.0(5)	& 746(24) 	&	594(3)\\
    \hline
    \end{tabular}
  \end{center}
\end{table}

\section{Discussion}
The intensity patterns observed in these data suggest that the these mirrors would be favorable for trapping of small clouds of ultra-cold atoms, or even single atoms \cite{Grangier:Nature:2001}.  For a total power of 1~W in the illuminating beam we estimate a peak intensity at the focus of $\sim10^9$~W/m$^2$; with a dipole trapping wavelength of $\lambda=852$~nm this corresponds to a trap depth of 370~$\mu$K for rubidium atoms \cite{Grimm:AdvAtMolOptPhys:2000}. The trap depth would scale at least with the square of the aperture of the mirrors (the power captured scales as the aperture diameter squared, while the peak intensity increases with the numerical aperture) meaning that deeper traps could readily be designed. Two dimensional arrays of single trapped atoms could be formed from one laser beam illuminating a congruent 2D array of mirrors. A feature of this array is that the illumination of individual mirrors would then provide single atom addressability, but on the length scale of the mirror aperture, which is inherently larger than the single atom focus. 

\begin{figure}[b!]
\centerline{\includegraphics[width=10cm]{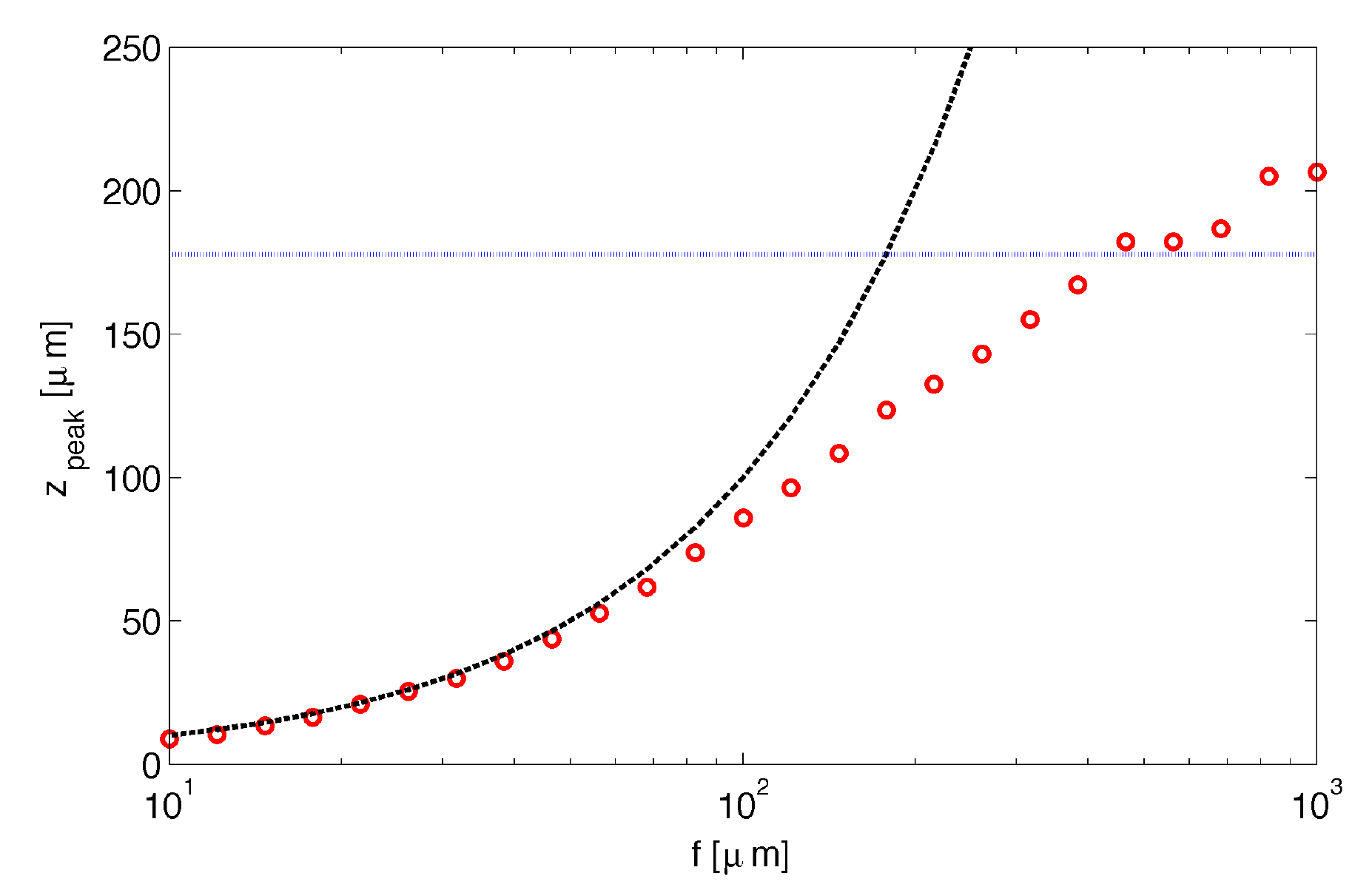}}
\caption{Comparison of angular spectrum simulations (red circles) and Gaussian optics (black dashed line) for the position of the peak intensity as a function of focal length for a mirror of aperture diameter $20~\mu$m. The blue dotted line corresponds to the position of peak intensity, the spot of Arago, after an open aperture of diameter equal to that of the simulated mirror.}
\label{fig:focal_vs_simulation}
\end{figure}

Although we have shown, in Fig.~\ref{fig:data_vs_simulation}, that the angular spectrum method yields agreement with the measured data in the regime probed here, further simulations show that focusing power of parabolic micro-mirrors is limited at longer focal lengths. In Fig.~\ref{fig:focal_vs_simulation} the position of peak intensity as a function of mirror design focal length shows that diffraction from the aperture edge begins to dominate at focal lengths significantly longer than the aperture radius. In the limit of large focal length the location of the peak intensity tends towards that predicted for diffraction from the edges; this is shown  in Fig.~\ref{fig:focal_vs_simulation} by the dotted line which is found from simulating diffraction from an aperture of radius equal the that of the mirrors. 

For the shortest focal length measured in this work, $f=24~\mu$m, we derive a numerical aperture of $N\!A_{\rm water}=0.95$ for operation in water. As shown in \cite{Salathe:OptExpr:2007} by Merenda {\it et al.}, stable 3D trapping is to be expected in this regime. Further work will be conducted to demonstrate optical tweezing using these mirrors.


\section{Conclusions}
Construction and diffraction limited focusing from parabolic micromirrors with radii of $10~\mu$m ($17\lambda$)  has been demonstrated. Simulations performed using the angular spectrum method show clear agreement with data and reproduce detailed features evident in the measured 3D intensity field. The results offer promise for use of these micromirrors for experiments in optical trapping in both biological and atomic-physics experiments.


\section*{Acknowledgments}
The authors are extremely grateful to Alastair Sinclair for initial project design and for discussions, to Jonathan Pritchard for help with simulations, and to John Harris for assistance. PFG acknowledges the generous support of the Royal Society of Edinburgh. 
\end{document}